\documentclass[prb,showpacs,preprint,nofloats,endfloats]{revtex4}

\usepackage{amsmath,amssymb,epsfig,epsf,psfig}

\begin{document}

\title
{\bf Pseudogaps in Strongly Correlated Metals:
A generalized dynamical mean-field theory approach}
\author{M.V.Sadovskii$^1$, I.A.Nekrasov$^{1,2}$, E.Z.Kuchinskii$^1$,
Th.Pruschke$^3$, V.I.Anisimov$^2$}

\affiliation
{$^1$Institute for Electrophysics, Russian Academy of Sciences,
Ekaterinburg, 620016, Russia\\
$^2$Institute for Metal Physics, Russian Academy of Sciences, 
Ekaterinburg, 620219, Russia\\
$^3$Institut f\"ur Theoretische Physik, Universit\"at G\"ottingen, Germany
}

\begin{abstract}
We generalize the dynamical--mean field (DMFT) approximation by including
into the DMFT equations some length scale  $\xi$ via a momentum dependent 
``external'' 
self--energy $\Sigma_{\bf k}$.  This external self--energy describes non-local
dynamical correlations induced by short--ranged collective SDW--like 
antiferromagnetic spin (or CDW--like charge) fluctuations.  At high enough 
temperatures these fluctuations can be viewed as a quenched Gaussian random 
field with finite correlation length. This generalized DMFT+$\Sigma_{\bf k}$ 
approach is used for the numerical solution of the weakly doped one--band 
Hubbard model with repulsive Coulomb interaction on a square lattice with 
nearest and next nearest neighbour hopping.  The effective single impurity 
problem in this generalized DMFT+$\Sigma_{\bf k}$ is solved by numerical 
renormalization group (NRG).  Both types of strongly correlated metals, 
namely (i) doped Mott insulator and (ii) the case of bandwidth $W\lesssim U$ 
($U$ --- value of local Coulomb interaction) are considered. Densities of 
states, spectral functions and ARPES spectra calculated within 
DMFT+$\Sigma_{\bf k}$ show a pseudogap formation near the Fermi level of the 
quasiparticle band.  
\end{abstract}


\maketitle

\newpage

\section{Introduction}

Among the numerous anomalies of the normal phase of high--temperature
superconductors the observation of a pseudogap
in the electronic spectrum of underdoped copper oxides \cite{Tim,MS}
is especially interesting.  
Despite continuing discussions on the nature of
the pseudogap, the preferable ``scenario'' for its formation 
is most likely based on the model of strong scattering of the charge
carriers by short--ranged antiferromagnetic (AFM, SDW) spin fluctuations
\cite{MS,Pines}. In momentum representation this scattering transfers 
momenta of the order of ${\bf Q}=(\frac{\pi}{a},\frac{\pi}{a})$ 
($a$ --- lattice constant of two dimensional lattice). 
This leads to the formation of structures in the one-particle spectrum, 
which are precursors of the changes in the spectra due
to long--range AFM order (period doubling).
As a result we obtain non--Fermi liquid like behaviour (dielectrization)
of the spectral density in the vicinity of the so called ``hot-spots'' on the
Fermi surface, appearing at intersections of the Fermi surface 
with antiferromagnetic Brillouin zone boundary (umklapp surface) \cite{MS}.

Within this spin--fluctuation scenario a simplified model of the pseudogap 
state was studied \cite{MS,Sch,KS} under the assumption that the scattering
by dynamic
spin fluctuations can be reduced for high enough temperatures
to a static Gaussian random field (quenched disorder) of pseudogap fluctuations.
These fluctuations are defined by a characteristic scattering vector from the 
vicinity of ${\bf Q}$,  with a width 
determined by the inverse correlation length of short--range
order $\kappa=\xi^{-1}$,  and by appropriate energy
scale $\Delta$ (typically of the order of crossover temperature $T^*$ to 
the pseudogap state \cite{MS}). 

Undoped cuprates are antiferromagnetic Mott insulators with
$U\gg W$ ($U$ --- value of local Coulomb interaction, $W$ --- bandwidth of
non--interacting band), so that
correlation effects are very important.
It is thus clear that the electronic properties of underdoped 
(and probably also optimally doped) cuprates are governed by strong electronic 
correlations too, so that these systems are typical strongly correlated
metals. Two types of correlated metals can be distinguished:
(i) the doped Mott insulator and (ii) the bandwidth controlled 
correlated metal $W\approx U$. Both types will be considered in this paper.

A state of the art tool to describe such correlated  systems
is the dynamical mean--field theory (DMFT)
\cite{MetzVoll89,vollha93,pruschke,georges96,PT}.
The characteristic features of correlated systems within the DMFT
are the formation of incoherent structures, the so-called Hubbard bands,
split by the Coulomb interaction $U$, and a
quasiparticle (conduction) band near the Fermi level dynamically 
generated by the local correlations 
\cite{MetzVoll89,vollha93,pruschke,georges96,PT}.

Unfortunately, the DMFT is not useful to the study the
``antiferromagnetic'' scenario of pseudogap formation in strongly
correlated metals. This is due to the basic approximation of the DMFT, which
amounts to the complete neglect of non-local dynamical correlation effects.

Besides the extended DMFT \cite{Si96}, which locally includes coupling
to nonlocal dynamical fluctuations, a straightforward way to extend the DMFT
are the so-called cluster mean-field theories \cite{TMrmp}. 
Two variants of this approach are the dynamical cluster approximation (DCA)\cite{TMrmp} and
cellular DMFT (CDMFT)\cite{KSPB}. 
In particular the DCA
has been applied to study the low-energy properties of
the Hubbard model, systematically including short- to medium ranged
nonlocal correlations.
Both improve on the cluster perturbation theory
(CPT) \cite{Gross94,Senechal00}, a
first attempts to use finite-size calculations to
obtain approximate results for the thermodynamic limit.

However, these approaches have certain drawbacks from both technical
and interpretation points of view.  
First, the effective quantum single
impurity problem becomes rather complex. Thus, most computational
methods available for the DMFT can be applied for the smallest
clusters only \cite{TMrmp,Kyung05,CivKot}, i.e.\ include
nearest-neighbor fluctuations only. For medium- to long-ranged
correlations one is currently restricted to Quantum Monte-Carlo\cite{QMC}. Since for cluster
problems again a sign problem arises, one is restricted to relatively
small values of the local Coulomb interaction and high temperatures. 
Second, the interpretation of electronic structures found has to be based on
reliable input from other, typically approximate, complementary techniques.

The aim of the present paper is to propose such a novel approach, which
on the one hand retains the single-impurity description of the DMFT, viz a proper
account for {\em local} correlations and the possibility to use very
efficient impurity solvers like NRG\cite{NRG,BPH}; on the other hand, we include
non-local correlations on a non-perturbative model basis, which allows to control
characteristic scales and also types of non-local fluctuations. This
latter point allows for a systematical study of the influence of
non-local fluctuations on the electronic properties and in particular
provide valuable hints on physical origin and possible interpretation
of results found in e.g.\ more refined theoretical approaches.

The paper is organized as follows:
In section \ref{leng_intro} we present a derivation of the
self--consistent generalization we call DMFT+$\Sigma_{\bf k}$ 
which includes short-ranged dynamical correlations to some extent.
Section \ref{kself} describes the construction of the
{\bf k}--dependent self--energy, and some 
computational details are presented in section \ref{compdet}.  
Results and a discussion are given in the sections \ref{results}.
Then the paper is ended with summary section~\ref{concl}
together with overview of related recent approaches and results
on pseudogap issue.

\section{Introducing length scale into DMFT: DMFT+$\Sigma_{\bf k}$ approach}
\label{leng_intro}

The basic shortcoming of the traditional DMFT approach
\cite{MetzVoll89,vollha93,pruschke,georges96,PT}
is the neglect of momentum dependence of the electron self--energy.
This approximation in principle allows for an exact solution of correlated
electron systems fully preserving the local part of the dynamics introduced
by electronic correlations.
To include non--local effects, while remaining within the usual ``single impurity
analogy'', we propose the following procedure. To be definite, let us consider
a standard one-band Hubbard model from now on. The extension to multi-orbital 
or multi-band models is straightforward.
The major assumption of our approach is that the lattice
and Matsubara ``time'' Fourier transformed of the single-particle Green function 
can be written as:
\begin{equation}
G_{\bf k}(i\omega)=\frac{1}{i\omega+\mu-\varepsilon({\bf k})-\Sigma(i\omega)
-\Sigma_{\bf k}(i\omega)},\qquad \omega=\pi T(2n+1),
\label{Gk}
\end{equation}
where $\Sigma(i\omega)$ is the {\em local} contribution to self--energy,
surviving in the DMFT, while $\Sigma_{\bf k}(i\omega)$
is some momentum dependent part. We suppose that
this last contribution is due to either electron interactions with some
``additional'' collective modes or order parameter fluctuations, or may be
due to similar non--local contributions within the Hubbard model itself. 

To avoid possible confusion we must stress that $\Sigma_{\bf k}(i\omega)$
can in principle also contain  local (momentum independent) contributions, which obviously
{\em vanish} in the limit of infinite dimensionality $d\to\infty$ and are
not taken into account within DMFT. Due to this fact there is no double
counting of diagrams within our approach to the Hubbard model.
This question does not arise at all if we consider $\Sigma_{\bf k}(i\omega)$
appearing due to some ``additional'' interaction. More important is that the 
assumed additive form of the self--energy $\Sigma(i\omega)+\Sigma_{\bf k}(i\omega)$
implicitly corresponds to neglect of possible interference
of these local (DMFT) and non--local contributions.
Furthermore, both contributions to the total self-energy
$\Sigma(i\omega)+\Sigma_{\bf k}(i\omega)$
individually obeye causality by construction. Thus, the sum
and finally the propagator~(\ref{Gk}) constructed from it
are causal, too. 

The self--consistency equations of our generalized DMFT+$\Sigma_{\bf k}$ 
approach are formulated as follows:
\begin{enumerate}
\item{Start with some initial guess of {\em local} self--energy
$\Sigma(i\omega)$, e.g. $\Sigma(i\omega)=0$}.  
\item{Construct $\Sigma_{\bf k}(i\omega)$ within some (approximate) scheme, 
taking into account interactions with collective modes or order parameter
fluctuations which in general can depend on $\Sigma(i\omega)$
and $\mu$.} 
\item{Calculate the local Green function  
\begin{equation}
G_{ii}(i\omega)=\frac{1}{N}\sum_{\bf k}\frac{1}{i\omega+\mu
-\varepsilon({\bf k})-\Sigma(i\omega)-\Sigma_{\bf k}(i\omega)}.
\label{Gloc}
\end{equation}
}
\item{Define the ``Weiss field''
\begin{equation}
{\cal G}^{-1}_0(i\omega)=\Sigma(i\omega)+G^{-1}_{ii}(i\omega).
\label{Wss}
\end{equation}
}
\item{Using some ``impurity solver'' to calculate the single-particle Green 
function for the effective single Anderson impurity problem, defined by 
Grassmanian integral 
\begin{equation}
G_{d}(\tau-\tau')=\frac{1}{Z_{\text{eff}}}
\int Dc^+_{i\sigma}Dc_{i\sigma}
c_{i\sigma}(\tau)c^+_{i\sigma}(\tau')\exp(-S_{\text{eff}})
\label{AndImp}
\end{equation}
with effective action for a fixed site (``single impurity'') $i$
\begin{equation}
S_{\text{eff}}=-\int_{0}^{\beta}d\tau_1\int_{0}^{\beta}
d\tau_2c_{i\sigma}(\tau_1){\cal G}^{-1}_0(\tau_1-\tau_2)c^+_{i\sigma}(\tau_2)
+\int_{0}^{\beta}d\tau Un_{i\uparrow}(\tau)n_{i\downarrow}(\tau)\;\;,
\label{Seff}
\end{equation}
$Z_{\text{eff}}=\int Dc^+_{i\sigma}Dc_{i\sigma}\exp(-S_{\text{eff}})$, and
$\beta=T^{-1}$. This step produces a {\em new} set of values 
$G^{-1}_{d}(i\omega)$.}
\item{Define a {\em new} local self--energy
\begin{equation}
\Sigma(i\omega)={\cal G}^{-1}_0(i\omega)-
G^{-1}_{d}(i\omega).
\label{StS}
\end{equation}
}
\item{Using this self--energy as ``initial'' one in step 1, continue the 
procedure until (and if) convergency is reached to obtain
\begin{equation}
G_{ii}(i\omega)=G_{d}(i\omega).
\label{G00}
\end{equation}
}
\end{enumerate}
Eventually, we get the desired Green function in the form of (\ref{Gk}),
where $\Sigma(i\omega)$ and $\Sigma_{\bf k}(i\omega)$ are those appearing
at the end of our iteration procedure.
A more detailed derivation of this scheme within a diagrammatic approach 
is  given in the Appendix \ref{A}.

\section{Construction of {\bf k}--dependent self--energy}
\label{kself}
For the momentum dependent part of the single-particle self--energy we 
concentrate on the effects of scattering of electrons from collective 
short-range SDW--like antiferromagnetic spin (or CDW--like charge) 
fluctuations.
To calculate $\Sigma_{\bf k}(i\omega)$ for an electron moving in the quenched
random field of (static) Gaussian spin (or charge) fluctuations with dominant
scattering momentum transfers from the vicinity of some characteristic
vector ${\bf Q}$ (``hot-spots'' model \cite{MS}), 
we use a slightly generalized version of the recursion procedure 
proposed in Refs.~\cite{Sch,KS,MS79} which takes into account {\em all} 
Feynman diagrams describing the scattering of electrons by this random field.  
This becomes possible due to a remarkable property of our simplified version
of ``hot-spots'' model that  under certain conditions
{\em the contribution of an arbitrary diagram 
with intersecting interaction lines is actually equal to the contribution of 
some diagram of the same order without intersections of these lines} 
\cite{KS,MS79}. 
Thus, in fact we can limit ourselves to consideration of only  
diagrams without intersecting interaction lines, taking the contribution of 
diagrams with intersections into account with the help of additional 
combinatorial factors, which are attributed to ``initial'' vertices or just 
interaction lines \cite{MS79}. 
As a result we obtain the following recursion 
relation (continuous fraction representation \cite{MS79}):
\begin{equation}
\Sigma_{n}(i\omega{\bf k})=\Delta^2\frac{s(n)}
{i\omega+\mu-\Sigma(i\omega)
-\varepsilon_n({\bf k})+inv_n\kappa-\Sigma_{n+1}(i\omega,{\bf k})}\;\;. 
\label{rec}
\end{equation}
Term $\Sigma_{n}(i\omega,{\bf k})$ of recurring sequence contains
all contributions of diagrams with the number of interaction lines $\geq n$.
Then  
\begin{equation}
\Sigma_{\bf k}(i\omega)=\Sigma_{n=1}(i\omega,{\bf k})
\label{Sk}
\end{equation}
is the sum of all diagrammatic contributions up to $2n$-th order.
Since the convergence of this recursion procedure for 
$\Sigma_{n}(i\omega,{\bf k})$ is rather fast, 
one can take contribution for large enough n equal to zero
and doing recursion backwards to $n=1$ get desired physical self--energy 
\cite{KS}.
 
The quantity $\Delta$ characterizes the energy scale and
$\kappa=\xi^{-1}$ is the inverse correlation length of short--range
SDW (CDW) fluctuations, $\varepsilon_n({\bf k})=\varepsilon({\bf k+Q})$ and 
$v_n=|v_{\bf k+Q}^{x}|+|v_{\bf k+Q}^{y}|$ 
for odd $n$ while $\varepsilon_n({\bf k})=\varepsilon({\bf k})$ and $v_{n}=
|v_{\bf k}^x|+|v_{\bf k}^{y}|$ for even $n$. The velocity projections
$v_{\bf k}^{x}$ and $v_{\bf k}^{y}$ are determined by usual momentum derivatives
of the ``bare'' electronic energy dispersion $\varepsilon({\bf k})$. Finally,
$s(n)$ represents a combinatorial factor with
\begin{equation}
s(n)=n
\label{vcomm}
\end{equation}
for the case of commensurate charge (CDW-type) fluctuations with
${\bf Q}=(\pi/a,\pi/a)$ \cite{MS79}. 
For incommensurate CDW fluctuations\cite{MS79} (when ${\bf Q}$ is not
``locked'' to the period of inverse lattice) one finds
\begin{equation} 
s(n)=\left\{\begin{array}{cc}
\frac{n+1}{2} & \mbox{for odd $n$} \\
\frac{n}{2} & \mbox{for even $n$}.
\end{array} \right.
\label{vinc}
\end{equation}
If we want to take into account the (Heisenberg) spin structure of interaction 
with spin fluctuations in  ``nearly antiferromagnetic Fermi--liquid'' 
(spin--fermion (SF) model of Ref.~\cite{Sch}, SDW-type fluctuations),
the combinatorics of diagrams becomes more complicated.
Spin--conserving scattering processes obey commensurate combinatorics,
while spin--flip scattering is described by diagrams of incommensurate
type (``charged'' random field in terms of Ref.~\cite{Sch}). In this model
the recursion relation for the single-particle Green function is again given by
(\ref{rec}), but the combinatorial factor $s(n)$ now acquires the following 
form \cite{Sch}:
\begin{equation} 
s(n)=\left\{\begin{array}{cc}
\frac{n+2}{3} & \mbox{for odd $n$} \\
\frac{n}{3} & \mbox{for even $n$}.
\end{array} \right.
\label{vspin}
\end{equation}
Obviously, with this procedure we introduce an important length scale $\xi$ 
not present in standard DMFT. Physically this scale mimics the effect of 
short--range (SDW or CDW) correlations within fermionic ``bath'' surrounding 
the effective single Anderson impurity of the DMFT.
We expect that such a length-scale will lead to a competition
between local and non-local physics.

An important aspect of the theory is that both parameters $\Delta$ and 
$\xi$ can in principle be calculated from the microscopic model at hand. 
For example, using the two--particle selfconsistent approach of Ref.~\cite{VT} 
with the approximations introduced in Refs.~\cite{Sch,KS}, one can derive 
within the standard Hubbard model the following microscopic expression 
for $\Delta$:
\begin{eqnarray} 
\Delta^2=\frac{1}{4}U^2\frac{<n_{i\uparrow}n_{i\downarrow}>}
{<n_{i\uparrow}><n_{i\downarrow}>}[<n_{i\uparrow}>+<n_{i\downarrow}>
-2<n_{i\uparrow}n_{i\downarrow}>]=\nonumber\\
=U^2\frac{<n_{i\uparrow}n_{i\downarrow}>}{n^2}<(n_{i\uparrow}
-n_{i\downarrow})^2>
=\nonumber\\
=U^2\frac{<n_{i\uparrow}n_{i\downarrow}>}{n^2}\frac{1}{3}<{\vec S}_i^2>,
\label{DeltHubb}
\end{eqnarray}
where we consider only scattering from antiferromagnetic spin fluctuations.
The different local quantities -- spin fluctuation $<{\vec S}_i^2>$,  density
$n$ and  double occupancy $<n_{i\uparrow}n_{i\downarrow}>$ -- 
can easily be calculated within the standard DMFT \cite{georges96}.
A detailed derivation of (\ref{DeltHubb}) and computational
results for $\Delta$ obtained by DMFT using quantum Monte--Carlo (QMC)
to solve the effective single impurity problem are presented in Appendix \ref{B}.
A corresponding microscopic expressions for the correlation length 
$\xi$ can also be derived within the two--particle self--consistent
approach \cite{VT}. However, we expect those results for $\xi$ to be less 
reliable, because this approach is valid only for relatively small (or medium) 
values of $U/t$.
Thus, in the following we will consider both $\Delta$ and especially $\xi$
as some phenomenological parameters to be determined from experiments.

\section{Results and discussion}
\label{results}

\subsection{Computation details}
\label{compdet}
In the following, we want to discuss results for a standard one-band
Hubbard model on a square lattice. With nearest ($t$) and
next nearest ($t'$) neighbour hopping integrals the dispersion
then reads
\begin{equation}
\varepsilon({\bf k})=-2t(\cos k_xa+\cos k_ya)-4t'\cos k_xa\cos k_ya\;\;,
\label{spectr}
\end{equation}
where $a$ is the lattice constant.
The correlations are introduced by a repulsive local two-particle 
interaction $U$.
We choose as energy scale the nearest neighbour hopping integral $t$
and as length scale the lattice constant $a$.

For a square lattice the bare bandwidth is $W=8t$.
To study a strongly correlated metallic state obtained as doped Mott insulator
we use $U=40t$ as value for the Coulomb interaction and a filling $n=0.8$ (hole 
doping).
The particular choice of the latter value for  $U$ is motivated
by two aspects. First, this value of
$U$ leads to an insulating DMFT+$\Sigma_{\bf k}$ solution at
half-filling.
Second, estimations of $U$ for stoichiometric La$_2$CuO$_4$
(high-T$_C$ prototype compound) based on
constrained LDA\cite{Gunnarsson89} calculations typically give
$U$ of the order of 10~eV\cite{LSCO_U_value},
which corresponds to 40t to our choice of parameters.
The correlated metal in the case of $W\gtrsim U$ 
is realized via $U=4t$ -- a value used in various theoretical papers
discussing the pseudogap state -- and two fillings: half-filling ($n=1.0$) and $n=0.8$ 
(hole doping). As typical values for $\Delta$ we choose $\Delta=t$ and 
$\Delta=2t$ (actually as approximate limiting values --- cf. Appendix~\ref{B})
and for the correlation length $\xi=2a$ and $\xi=10a$ (motivated mainly by 
experimental data for cuprates~\cite{MS,Sch}).

The DMFT maps the lattice problem onto an effective, 
self--consistent single impurity defined by Eqs. (\ref{AndImp})-(\ref{Seff}).  
In our work we employ as ``impurity solvers'' two 
reliable numerically exact methods --- quantum Monte--Carlo (QMC)~\cite{QMC} 
and numerical renormalization group (NRG) \cite{NRG,BPH}.
Calculations were done for the case $t'=0$ and 
$t'/t$=-0.4 (more or less typical for cuprates)
at two different temperatures $T=0.088t$ and $T=0.356t$ (for NRG 
computations)~\footnote{
Discretization parameter $\Lambda$=2, number of
low energy states after truncation 1000, cut-off near Fermi energy 10$^{-6}$,
broadening parameter b=0.6.}.
QMC computations of double occupancies as functions of 
filling were done at temperatures $T=0.1t$ and $T=0.4t$ 
~\footnote{
Number warm-up sweeps 30000, number of QMC sweeps 200000,
number of imaginary time slices 40.}. 

Below we present results only for most typical dependences and parameters,
more data and figures can be found in Ref. \cite{cm05}.

\subsection{Generalized DMFT+$\Sigma_{\bf k}$ approach: densities of states}

\begin{figure}[htb]
\includegraphics[clip=true,width=0.8\textwidth]{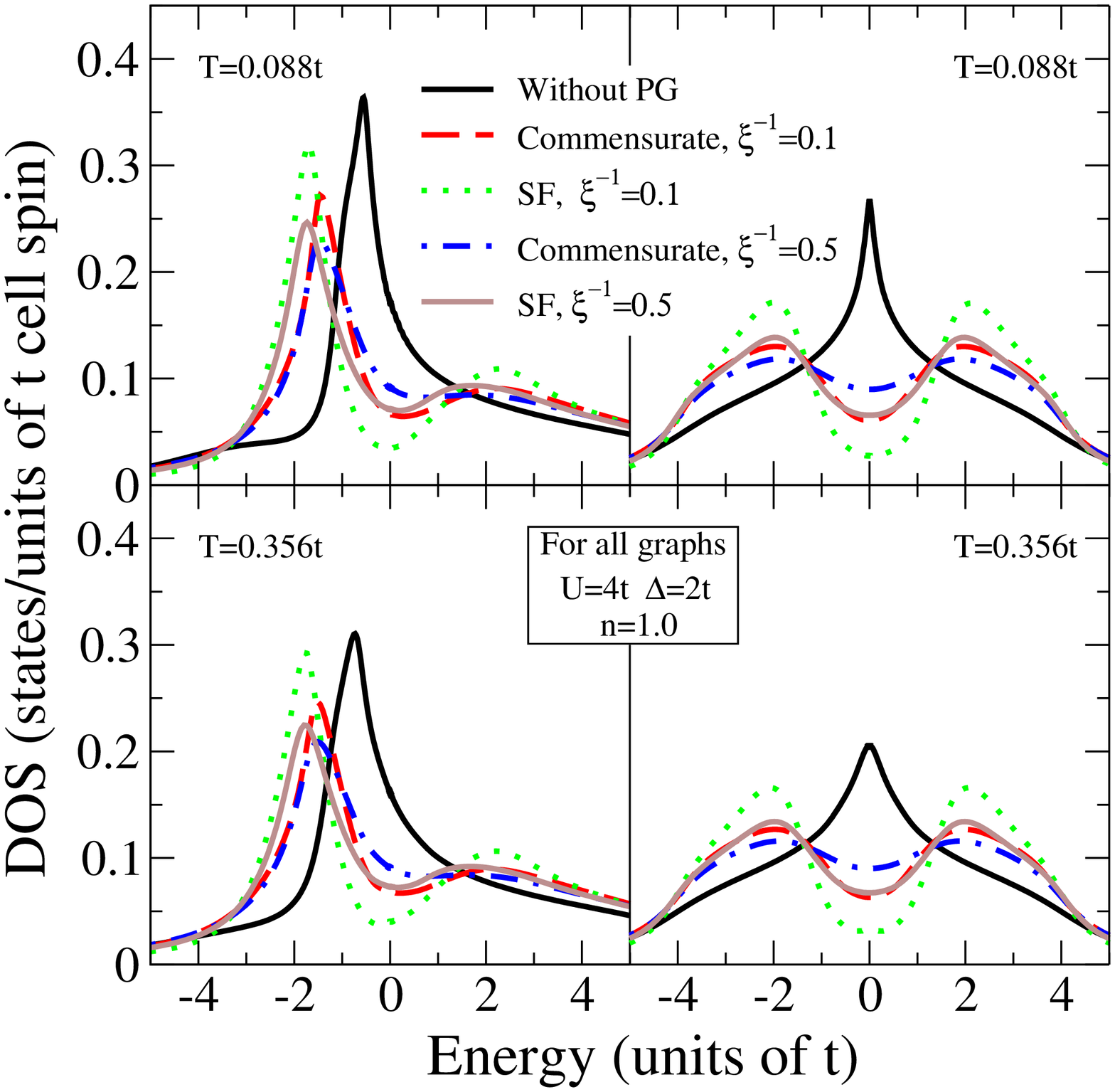}
\caption{(Color online) Comparison of DOS obtained from
DMFT(NRG)+$\Sigma_{\bf k}$ calculations for different
combinatorical factors (SF --- spin--fermion model, commensurate),
inverse correlation lengths
($\xi^{-1}$) in units of the lattice constant, temperatures ($T$) and 
the value of pseudogap potential $\Delta=2t$.
Left column corresponds to $t'/t=-0.4$, right column to $t'=0$.
In all graphs the Coulomb interaction is $U=4t$ and $n=1$.
The Fermi level corresponds to zero.}
\label{DOS_4t_n1}
\end{figure}

\begin{figure}[htb]
\includegraphics[clip=true,width=0.8\textwidth]{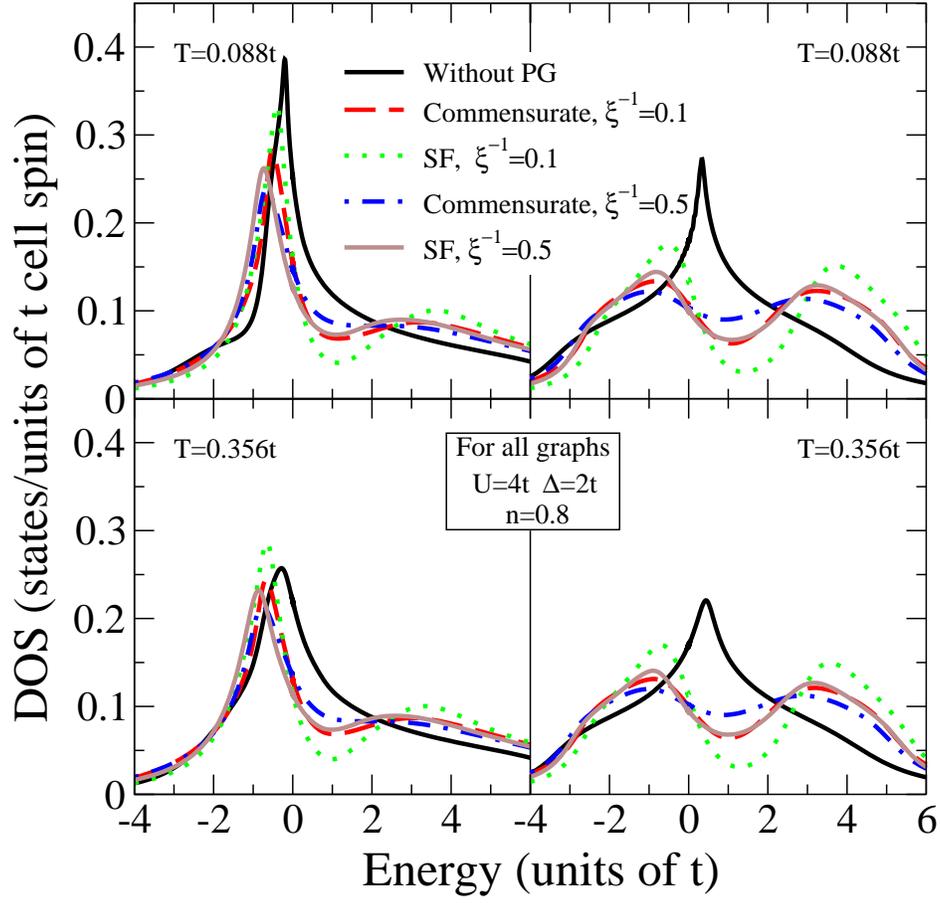}
\caption{(Color online) Comparison of DOS obtained from
DMFT(NRG)+$\Sigma_{\bf k}$ calculations for a filling $n=0.8$, other parameters
as in Fig.~\ref{DOS_4t_n1}.}
\label{DOS_4t_n08}
\end{figure}

Let us start the discussion of our results
obtained within our generalized DMFT+$\Sigma_{\bf k}$ approach
with the densities of states (DOS) for the case of small (relative
to bandwidth) Coulomb interaction $U=4t$ with and without pseudogap fluctuations.
As already discussed in the introduction, the characteristic feature of the 
strongly correlated metallic state
is the coexistence of lower and upper Hubbard bands split by the value of $U$
with a quasiparticle peak at the Fermi level.
Since at half--filling the bare DOS of the square lattice has a Van--Hove 
singularity at the Fermi level ($t'=0$) or close to it (in case of $t'/t=-0.4$) 
one cannot treat a peak on the Fermi level simply as a quasiparticle peak. 
In fact, there are two contributions to this peak from (i) 
the quasiparticle peak appearing in strongly correlated metals due to 
many-body effects and (ii) the smoothed Van--Hove singularity from the bare 
DOS~\footnote{We have checked that with increasing
of Coulomb repulsion the Van--Hove singularity {\em gradually} transforms into
quasiparticle peak for $U=(6\div 8)t$.}.  
In Figs.~\ref{DOS_4t_n1} and~\ref{DOS_4t_n08} we show the corresponding 
DMFT(NRG) DOS without pseudogap fluctuations as black lines for both
bare dispersions $t'/t=-0.4$ (left panels) and for $t'=0$ (right panels) for 
two different  temperatures $T=0.356t$ (lower panels) and $T=0.088t$ 
(upper panels) with fillings $n=1.0$ and $n=0.8$ respectively.
The remaining curves in Figs.~\ref{DOS_4t_n1} and~\ref{DOS_4t_n08} represent 
results for the DOS with non-local fluctuations switched on with
the fluctuation amplitude $\Delta=2t$.
For all sets of parameters one can see that the
introduction of non-local fluctuations into the calculation leads to the 
formation of pseudogap in the quasiparticle peak.

The behaviour of the pseudogaps in the DOS has some common features.
For example, for $t'$=0 at half--filling (Fig. \ref{DOS_4t_n1}, right column) 
we find that the pseudogap is most pronounced.
For $n=0.8$ (Fig. \ref{DOS_4t_n08}, right column) the picture is almost the 
same but slightly asymmetric. The width of the pseudogap (the distance between 
peaks closest to Fermi level) appears to be of the order of $\sim 2\Delta$ 
here. Decreasing the value of $\Delta$ from $2t$ to $t$ leads to a pseugogap 
that is correspondingly twice smaller and in addition more shallow
(see Ref.~\cite{cm05}).
When one uses the combinatorial factors corresponding to the
spin--fermion model (Eq.(\ref{vspin})), we find that the pseudogap becomes more 
pronounced than in the case of commensurate charge fluctuations (combinatorial 
factors of Eq. (\ref{vinc})). 
The influence of the correlation length $\xi$ can be seen is also as expected.
Changing from $\xi^{-1}=0.1$ to $\xi^{-1}=0.5$, i.e.\ decreasing the range
of the non-local fluctuations, slightly washes out the pseudogap.  
Also, increasing the temperature from $T=0.088t$ to $T=0.356t$ leads to a 
general broadening of the structures in the DOS.  
These observations remain at least qualitatively
valid for $t'/t=-0.4$ (Figs. \ref{DOS_4t_n1} and \ref{DOS_4t_n08}, left columns)
with an additional asymmetry due to the next-nearest neighbour hopping.
Noteworthy is however the fact that for $t'/t=-0.4$ and $\xi ^{-1}=0.5$ the 
pseudogap has almost disappeared for the temperatures studied here.  
Also very remarkable point is the similarity of the results 
obtained with the generalized DMFT+$\Sigma_{\bf k}$ approach with $U=4t$ 
(smaller than the  bandwidth $W$) to those obtained earlier without 
Hubbard--like Coulomb interactions \cite{Sch,KS}.

\begin{figure}[htb]
\includegraphics[clip=true,width=0.8\textwidth,angle=270]{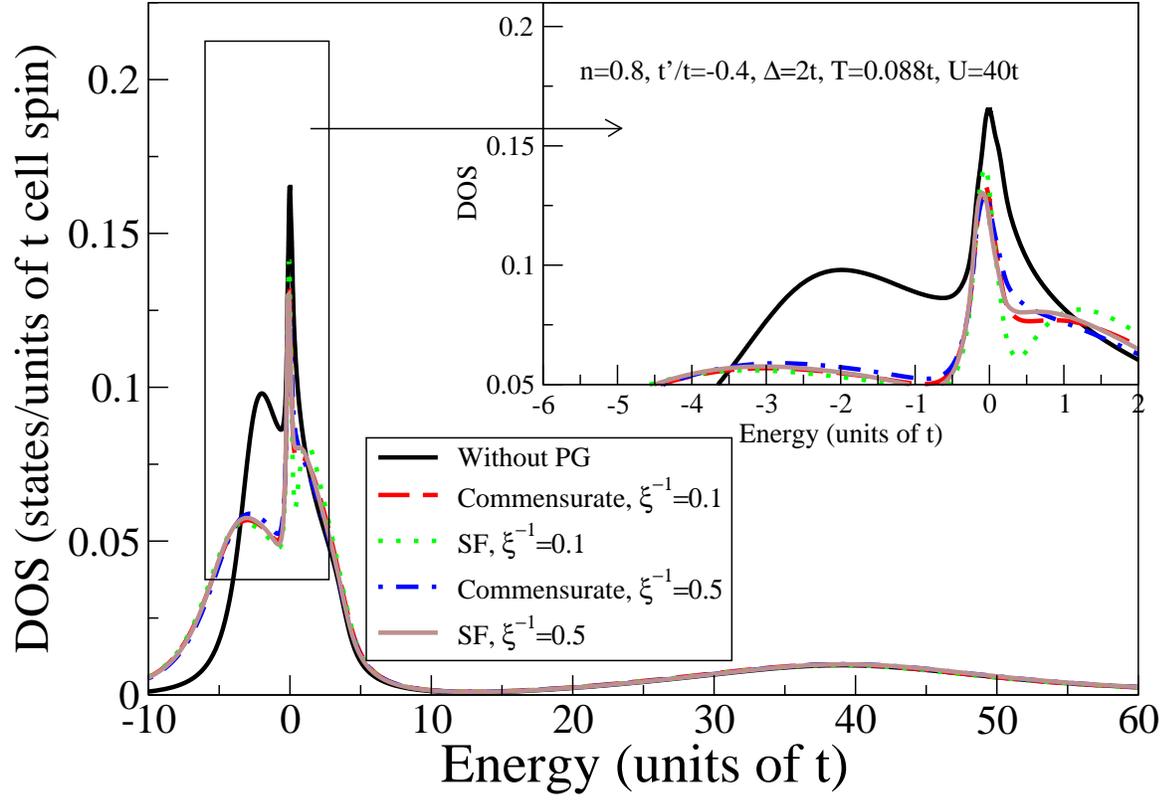}
\caption{(Color online) Comparison of DOS obtained from
DMFT(NRG)+$\Sigma_{\bf k}$ calculations for $t'/t=-0.4$, $T=0.088t$,
$U=40t$, $\Delta=2t$ and filling $n=0.8$.}
\label{dos_40t_04}
\end{figure}

Let us now consider the case of a doped Mott insulator.  The model parameters
are $t'/t=-0.4$ with filling 
$n=0.8$, but the Coulomb interaction strength is now set to $U=40t$. 
Characteristic features of the DOS 
for such a strongly correlated metal are a strong separation of lower and 
upper Hubbard bands and a Fermi level crossing by the lower Hubbard band (for 
non--half--filled case).  Without non-local fluctuations the quasi-particle 
peak is again formed at the Fermi level; but now the upper Hubbard band is 
far to the right and does not touch the quasiparticle peak (as it was for the 
case of small Coulomb interactions). DOS without non-local fluctuations 
are again presented as black lines in Fig.~\ref{dos_40t_04}.  Results for the 
case $t'=0$ are  presented elsewhere \cite{cm05}.

With rather strong non-local fluctuations $\Delta =2t$,
a pseudogap appears in the middle of quasiparticle peak. 
In addition we observe that
the lower Hubbard band is slightly broadened by fluctuation effects.
Qualitative behaviour of the pseudogap anomalies is again similar to those
described above for the case of $U=4t$, e.g.\ a decrease of $\xi$ makes 
the pseudogap less pronounced, reducing $\Delta$ from $\Delta =2t$ to 
$\Delta =t$ narrows of the pseudogap and also makes it more shallow etc.
(see Ref.~\cite{cm05}). 
Note that for the doped Mott--insulator we find that
the pseudogap is remarkably more pronounced for the SDW--like fluctuations
than for CDW--like fluctuations.

There are, however, obvious differences to the case with $U=4t$. 
For example, the width of the pseudogap appears to be much smaller than 
$2\Delta$, beeing of the order of $\Delta/2$ instead (see Fig.~\ref{dos_40t_04}).
This effect
we attribute to the fact that the quasiparticle peak itself 
is actually strongly narrowed now by local correlations. 

\subsection{Generalized DMFT+$\Sigma_{\bf k}$ approach:
spectral functions $A(\omega,{\bf k})$}

\begin{figure}[htb]
\includegraphics[clip=true,width=0.6\textwidth,angle=270]{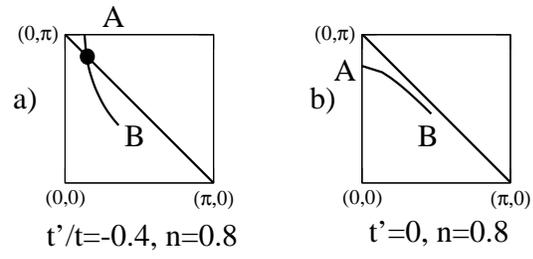}
\caption{1/8-th of the bare Fermi surfaces for
the occupancy $n=0.8$ and different combinations ($t,t'$)
used for the calculation of spectral functions $A({\bf k},\omega)$.
The diagonal line corresponds to the umklapp surface. The full circle marks 
the so-called ``hot--spot''.}
\label{FS_shapes}
\end{figure}

In the previous subsections we discussed the densities of states obtained
self--consistently by the DMFT+$\Sigma_{\bf k}$ approach. Once we get a
self--consistent solution of the DMFT+$\Sigma_{\bf k}$ equations with
non-local fluctuations we can of course also compute the spectral functions
$A(\omega,{\bf k})$
\begin{equation}
A(\omega,{\bf k})=-\frac{1}{\pi}{\rm Im}\frac{1}{\omega+\mu
-\varepsilon({\bf k})-\Sigma(\omega)-\Sigma_{\bf k}(\omega)},
\label{specf}
\end{equation}
where self--energy $\Sigma(\omega)$ and chemical potential $\mu$
are calculated self--consistently as described in Sec. \ref{leng_intro}. 
To plot $A(\omega,{\bf k})$ we choose ${\bf k}$--points along the
``bare'' Fermi surfaces for different types of
lattice spectra and filling $n=0.8$. In Fig. \ref{FS_shapes}
one can see corresponding shapes of these ``bare'' Fermi surfaces (presented 
are only 1/8-th of the Fermi surfaces within the first quadrant of the
first Brillouin zone).

\begin{figure}
\includegraphics[clip=true,width=0.8\textwidth]{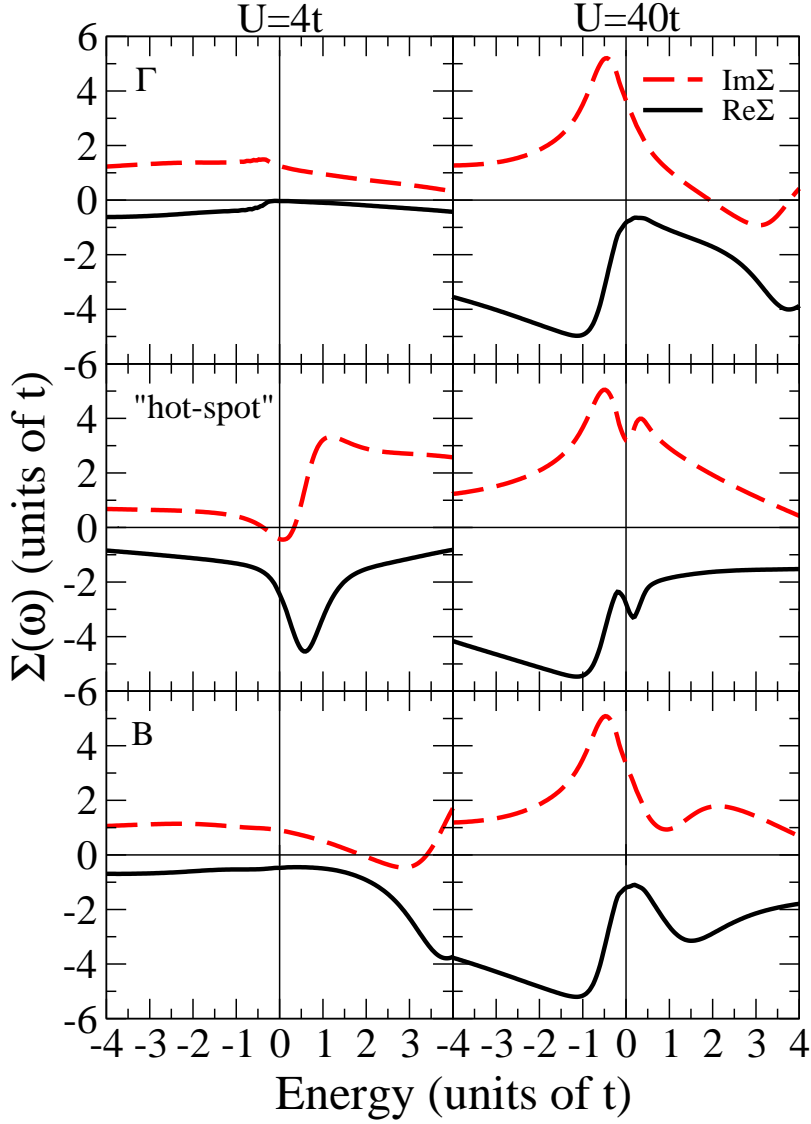}
\caption{(Color online) Real (dashed line) and imaginary (full line)
parts of the self-energy $\Sigma({\bf k},\omega)$
for $t/t'=-0.4$, $U=4t$ (left column)
and $U=40t$ (right column) for characteristic ${\bf k}$-points:
$\Gamma$, ``hot-spot'' (see Fig.~\ref{FS_shapes}) and ``cold-spot''
(point ``B'' in Fig.~\ref{FS_shapes}).
For all graphs the filling is $n=0.8$, temperature $T=0.088t$, inverse correlation
length $\xi^{-1}=0.1$, value of pseudogap potential $\Delta=2t$,
and SF combinatorics.}
\label{sigma_n08}
\end{figure}

A first natural quantity to inspect is the self-energy
$\Sigma({\bf k},\omega+i\delta)$, shown in Fig.~\ref{sigma_n08}
for $t'/t=-0.4$, $n=0.8$ and $U=4t$ (left column) and $U=40t$ (right
column). As representative ${\bf k}$-points we chose the centre of the
first Brillouin zone ($\Gamma$), the ``hot-spot'' and ``cold-spot''
(point ``B'' in Fig.~\ref{FS_shapes}).
The results were obtained with NRG at a temperature $T=0.088t$.
The structures for $U=4t$ are rather broad, but reveal after a closer
inspection features similar to the case $U=40t$. For the latter, the behaviour at
$\Gamma$ and ``B'' is very different from the structures at the ``hot-spot''.
Namely, while for the former two ${\bf k}$-points
${\rm Im}\Sigma({\bf k},\omega+i\delta)$ shows a nice parabolic maximum at the
Fermi energy, the latter develops a minimum instead. Such a structure
in the self-energy will result in a rather evident (pseudo) gap
in the spectral function at this {\bf k}-point and weaker pseudogap 
behaviour in the DOS. Its appearance is obviously due to the presence of the
spin-fluctuations at the ``hot-spot''. Note that similar features have
been observed in numerically expensive cluster mean-field calculations
\cite{mpj03}, too, with an interpretation as spin-fluctuation induced
based on physical expectations. Our calculations, obtained at a
minimum numerical expense, indeed show, that
including short-ranged fluctuations will precisely produce these non
Fermi-liquid structures in the one-particle self-energy.
This behaviour is quite typical for the problem and was observed
by other groups using different methods\cite{Kyung05,Stanescu03,Katanin04,RM}.
In several works midgap peak in the pseudogap was obtained with
explanation of its origin by particular shape of the self-energy
close to the Fermi level\cite{Katanin04,Stanescu03,Barisic05}.

\begin{figure}[htb]
\includegraphics[clip=true,width=0.8\textwidth]{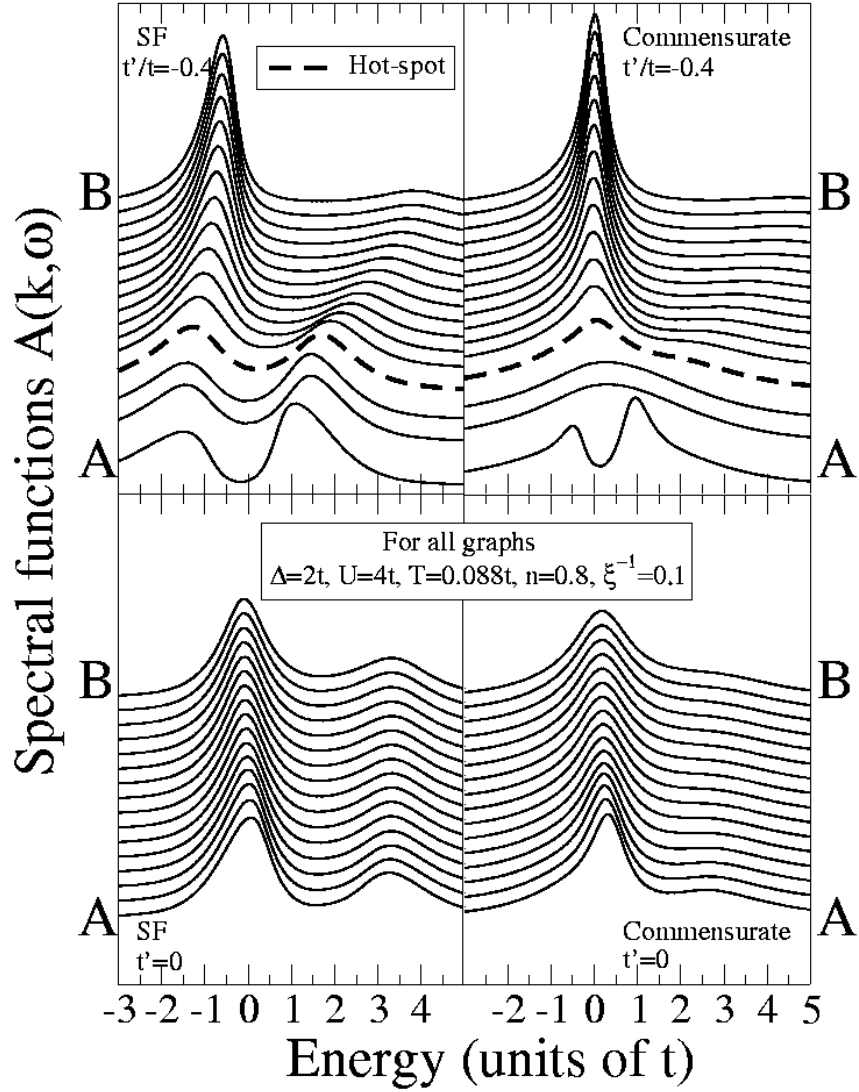}
\caption{Spectral functions $A({\bf k},\omega)$ 
obtained from the DMFT(NRG)+$\Sigma_{\bf k}$ calculations along the
directions shown in Fig. \ref{FS_shapes}. 
Model parameters were chosen as 
$U=4t$, $n=0.8$, $\Delta=2t$, $\xi^{-1}=0.1$ 
and temperature $T=0.088t$. 
The ``hot--spot'' {\bf k}-point is marked as fat dashed line.
The Fermi level corresponds to zero.}
\label{sf_U4t_n08}
\end{figure}

In the following we concentrate mainly on the case $U=4t$ and
filling $n=0.8$ (Fermi surface of Fig. \ref{FS_shapes}(a)). The
corresponding spectral functions $A(\omega,{\bf k})$ are depicted in
Fig.~\ref{sf_U4t_n08}. 
When $t'/t=-0.4$ (upper row), the spectral function close to the Brillouin zone 
diagonal (point B) has the typical Fermi--liquid behaviour, consisting of a 
rather sharp peak close to the Fermi level.
In the case of SDW--like fluctuations 
this peak is shifted down in energy by about $-0.5t$ (left upper corner).
In the vicinity of the ``hot--spot'' the shape of $A(\omega,{\bf k})$ is 
completely modified. Now $A(\omega,{\bf k})$ becomes double--peaked and
non--Fermi--liquid--like.  Directly at the ``hot--spot'', $A(\omega,{\bf k})$ 
for SDW--like  fluctuations has two equally intensive peaks situated 
symmetrically around the Fermi level and split from each other by $\sim 
1.5\Delta$ Refs.~\cite{Sch,KS}.  For commensurate CDW--like fluctuations the
spectral function in the 
``hot--spot'' region has one broad peak centred at the Fermi level 
with width $\sim \Delta$. Such a merging of the two peaks at the ``hot--spot'' 
for commensurate fluctuations was previously observed in Ref. \cite{KS}.
However close to point A  this type of fluctuations also produces a
double--peak structure in the spectral function.

Spectral functions for the case of
$U=4t$ at half--filling ($n=1$) 
and for $t'/t=-0.4$  are
similar to those just discussed for $n=0.8$. However, the
pseugogap is more pronounced in this case and remains open everywhere close 
to the umklapp surface for SDW fluctuations \cite{cm05}.

In the lower panel of Fig.~\ref{sf_U4t_n08} we show spectral functions
for 20\% hole doping ($n=0.8$) and the case of $t'=0$ (Fermi surface from 
Fig. \ref{FS_shapes}(b)). Since the Fermi surface now is close to the
umklapp surface, the pseudogap anomalies are rather strong and 
almost non--dispersive along the Fermi surface.  At half--filling
for $t'=0$ the Fermi surface actually coincides with umklapp surface
(in case of perfect ``nesting'' whole Fermi surface 
is the ``hot--region'').  
The spectral functions are now symmetric around the Fermi level.  
For SDW--like fluctuations there are two peaks 
split by $\sim 1.5\Delta$. Again, CDW--like fluctuations give just one 
peak centred at the Fermi level with width $\sim \Delta$.

\begin{figure}[htb]
\includegraphics[clip=true,width=0.8\textwidth]{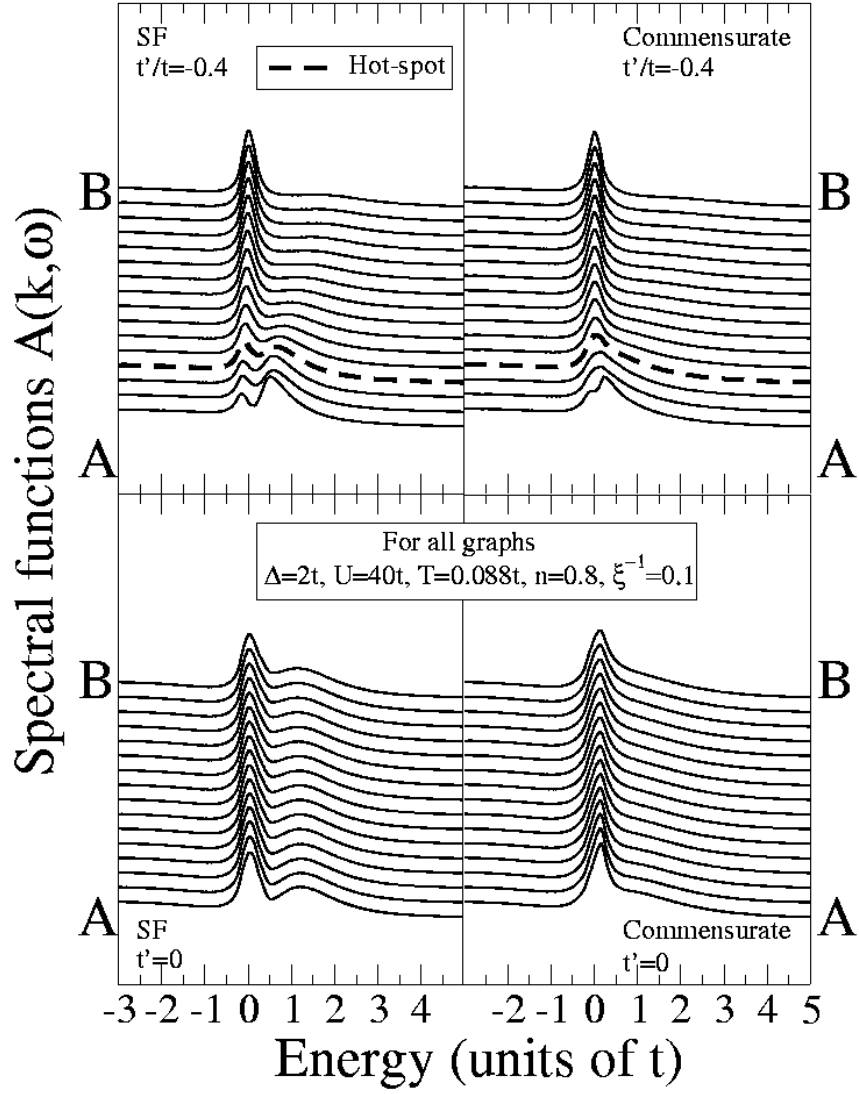}
\caption{Spectral functions $A({\bf k},\omega)$ 
obtained from the DMFT(NRG)+$\Sigma_{\bf k}$ calculations
for $U=40t$,
other parameters as in Fig.~\ref{sf_U4t_n08}.}
\label{sf_U40t_n08}
\end{figure}

For the case of a doped Mott insulator ($U=40t$, $n=0.8$), the spectral 
functions obtained by the DMFT+$\Sigma_{\bf k}$ approach are presented in 
Fig.~\ref{sf_U40t_n08}.
Qualitatively, the shapes of these spectral functions are similar to 
those shown in Fig.~\ref{sf_U4t_n08}. 
As was pointed out above, the strong Coulomb correlations lead to a
narrowing of the quasiparticle peak and a corresponding decrease of the  
pseudogap width.
As is evident from Fig.~\ref{sf_U40t_n08} the structures connected
    to the pseudogap are now
spread in an energy interval $\sim t$, while for $U=4t$ they are 
restricted to an interval $\sim 4t$ instead.
One should also note that in contrast to $U=4t$
the spectral functions are now about four times less intensive,
because part of the spectral weight is transferred to the upper Hubbard band located 
at about $40t$ and well separated from the quasiparticle peak now.

\begin{figure}[htb]
\includegraphics[clip=true,width=0.8\textwidth]{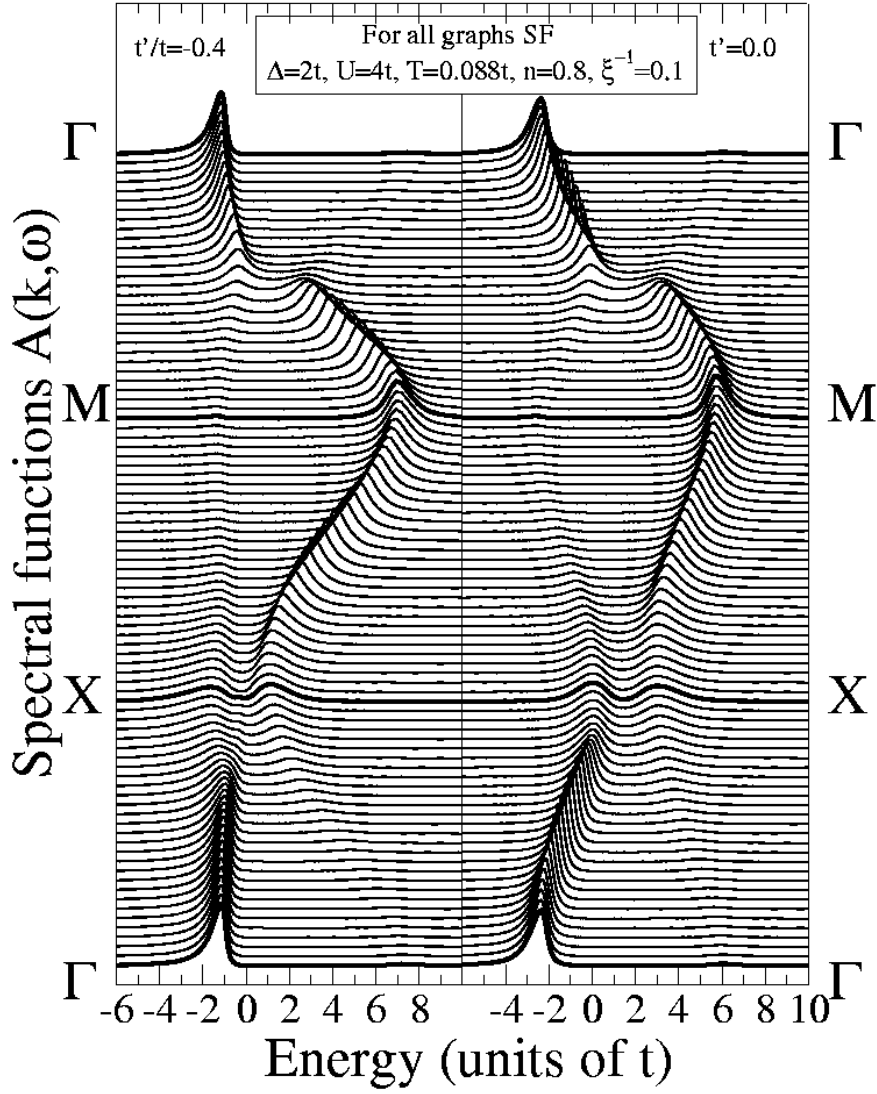}
\caption{Spectral functions $A({\bf k},\omega)$ obtained from the
DMFT(NRG)+$\Sigma_{\bf k}$ calculations
along high-symmetry directions of first Brillouin zone
$\Gamma(0,0)\!-\!\rm{X}(\pi,0)\!-\!\rm{M}(\pi,\pi)\!-\!\Gamma(0,0)$,
SF combinatorics (left row) and commensurate combinatorics (right column).
Other parameters are $U=4t$, $n=0.8$, $\Delta=2t$, $\xi^{-1}=0.1$ 
and temperature $T=0.088t$.
The Fermi level corresponds to zero.}
\label{n08_U4t_tri}
\end{figure}

Using another quite common choice of ${\bf k}$--points we can compute 
$A(\omega,{\bf k})$ along high--symmetry directions in the first 
Brillouin zone: 
$\Gamma(0,0)\!-\!\rm{X}(\pi,0)\!-\!\rm{M}(\pi,\pi)\!-\!\Gamma(0,0)$.  
The spectral functions for these 
${\bf k}$--points are collected in Fig.~ \ref{n08_U4t_tri}
for the case of SDW--like fluctuations.
Characteristic curves for doped Mott insulator are presented in Ref.\cite{cm05}.
For all sets of parameters one can 
see a characteristic double -- peak pseudogap structure close to the $X$ point. 
In the middle of $M\!-\!\Gamma$ direction (so called ``nodal'' point)
one can see the reminiscence of
AFM gap which has its biggest value here in case of perfect
antiferromagnetic ordering. Also in the nodal point ``kink''-like
behaviour is observed caused by interactions between correlated electrons 
with short--range pseudogap fluctuations.
A change of the filling leads mainly to a 
rigid shift of spectral functions with respect to the Fermi level. 

\begin{figure}
\includegraphics[clip=true,width=0.8\textwidth]{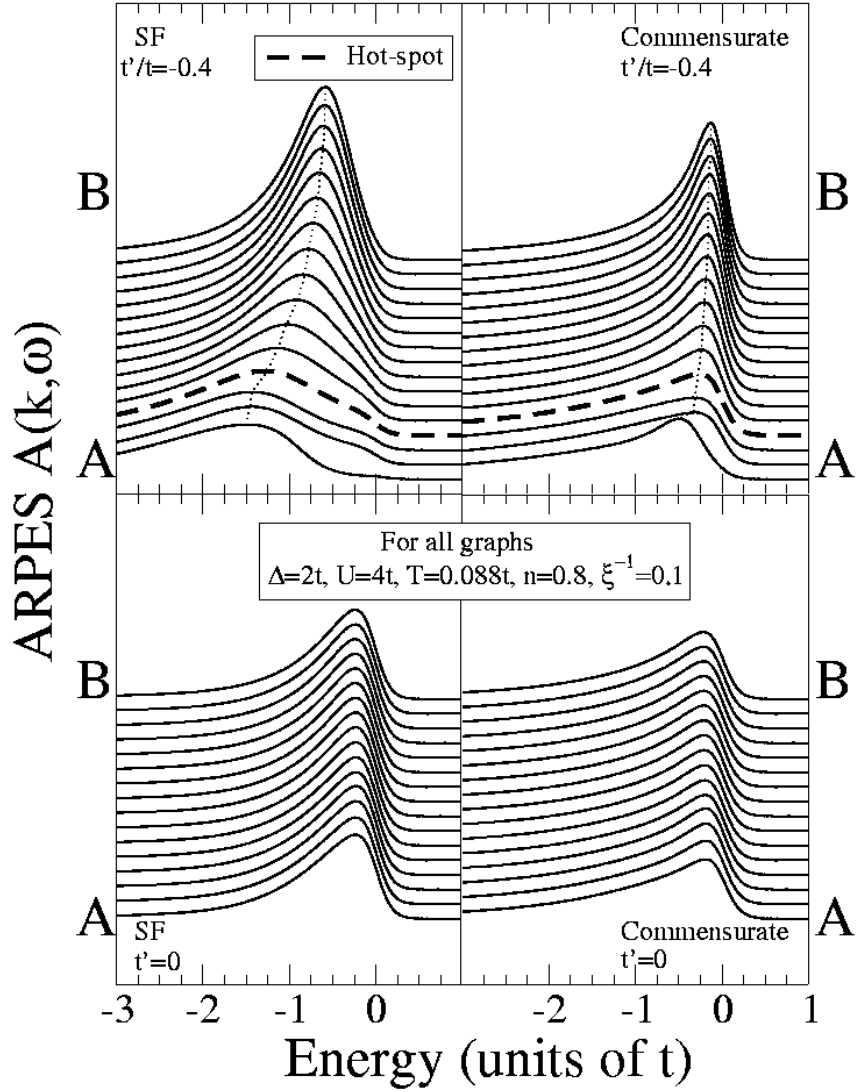}
\caption{ARPES spectra simulated by multiplication of the spectral functions
obtained from DMFT(NRG)+$\Sigma_{\bf k}$ calculations for $U=4t$ and $n=0.8$
Fig.~\ref{sf_U4t_n08} with Fermi function at $T=0.088t$ plotted 
along the lines in the first BZ as depicted by
Fig.~\ref{FS_shapes}. All other parameters are the same as in Fig.~\ref{sf_U4t_n08}.}
\label{arpes_U4t_n08}
\end{figure}

With the spectral functions we are now of course in a position to calculate
angle resolved photoemission spectra (ARPES), which is the most direct
experimental way to observe pseudogap in real compounds. 
For that purpose, we only
need to multiply our results for the spectral functions with the Fermi 
function at temperature $T=0.088t$. Typical example of the
resulting DMFT+$\Sigma_{\bf k}$ ARPES spectra are
presented in Fig.~\ref{arpes_U4t_n08}.
More figures of ARPES-like results obtained within the DMFT+$\Sigma_{\bf k}$ 
approach for a variety of parameters can be found in Ref.~\cite{cm05}.
One should note that for $t'/t=-0.4$ (upper panel of Fig.~\ref{arpes_U4t_n08}) 
as ${\bf k}$ goes from point ``A'' to point ``B'' the
peak situated slightly below the Fermi level changes its
position and moves down in energy.
Simultaneously it becomes more broad and less intensive.
The dotted line guides the motion of the peak maximum.
Also at the ``hot--spot'' and further to point ``B'' one can see some signs
of the double--peak structure.
Such behaviour of the peak in the ARPES is rather reminiscent of those observed
experimentally in underdoped cuprates \cite{MS,Sch,Kam}.

\section{Conclusion}
\label{concl}

To summarize, we propose a generalized DMFT+$\Sigma_{\bf k}$
approach, which is meant to take into account the important
effects of non--local correlations (in principle of any type)
in addition to the (essentially exact) treatment of local dynamical
correaltions by the DMFT.
In the standard DMFT the ``bath'' surrounding
the effective single Anderson impurity is spatially uniform
since the DMFT self-energy is only energy-dependent.
The main idea of our extension is to introduce non-local correlations
through the ``bath'', i.e.\ to make it spatially non-uniform,
while keeping standard DMFT self-consistency equations.
Such a generalization of the DMFT allows to supplement it
with a ${\bf k}$--dependent self--energy $\Sigma({\bf k},\omega)$.
It in turn opens the possibility to access the physics of low--dimensional 
strongly correlated systems, where different types of spatial fluctuations 
(e.g. of some order parameter) become important, in a non-perturbative way at
least with respect to the important local dynamical correlations. 
However, we must stress that our procedure in no way introduces any kind of
systematic $1/d$--expansion, being only a qualitative method to include
a length scale into DMFT. Nevertheless we believe that such a
technique can give valuable insight into the physical processes
leading to correlation induced $k$-dependent structures in single-particle properties.

In this work we model such effects for the two-dimensional
Hubbard model by incorporating into the ``bath''
scattering of fermions from non-local collective SDW--like
antiferromagnetic spin (or CDW--like charge) short-range fluctuations.
The corresponding ${\bf k}$--dependent self--energy $\Sigma({\bf k},\omega)$
is obtained from a non-perturbative iterative scheme \cite{Sch,KS}.
Such choice of the $\Sigma({\bf k},\omega)$ allows  to
address the problem of pseudogap formation in the
strongly correlated metallic state. We showed evidence that the pseudogap 
appears at the Fermi level within the quasiparticle peak, introducing a new
small energy scale of the order of psedogap potential value $\Delta$
in the DOS and more pronounced in spectral functions $A(\omega,{\bf k})$.
Let us stress, that our generalization of DMFT leads to non--trivial 
and in our opinion physically sensible ${\bf k}$--dependence of spectral 
functions. Is is significant that this particular choice of
$\Sigma({\bf k},\omega)$\cite{Sch,KS} does not cause difficulties
to ``double counting'' problem within our combined DMFT+$\Sigma_{\bf k}$ approach.
Also, the combination of diagrammatically correct techniques like
DMFT\cite{MetzVoll89,vollha93,pruschke,georges96,PT}
and the non-local self-energy ansatz of Refs.\cite{Sch,KS}
preserves the correct analytical properties of the combined self-energy
$\Sigma(i\omega)+\Sigma_{\bf k}(i\omega)$,
as well as of the corresponding one-electron propagator~(\ref{Gk}).

Of course our pseudogap observations are not entirely new.
Similar results about pseudogap formation in the 2d Hubbard
model were already obtained within cluster DMFT extensions, i.e.\ the
dynamical cluster approximation (DCA)\cite{TMrmp,mpj03} and
cellular DMFT (CDMFT) \cite{Kyung05,CivKot}, CPT
\cite{Gross94,Senechal00,Senechal05} and two interacting
Hubbard sites selfconsistently embedded in a bath \cite{Stanescu03}. However, these methods
have generic restrictions concerning the size of the cluster,
temperatures or filling accessible and, in case of the QMC, values of
the local Coulomb energy.
Recently, also the EDMFT was applied  to demonstrate 
pseudogap formation in the DOS due to dynamic Coulomb correlations\cite{HW}.
Note, however, that within the EDMFT there is no way to obtain a
${\bf k}$--dependence in spectral functions beyond that originating from the
bare electronic energy dispersion.
Important progress was also made  with weak coupling approaches for the
Hubbard model \cite{KyTr} and functional renormalization group \cite{Katanin04,RM}.
In several papers pseudogap formation was described in the framework
of the t-J model \cite{Prelovsek}.
A more general scheme for the inclusion of non--local corrections was also
formulated within the so called GW extension to the DMFT \cite{BAG,SKot}.

While at a first glance the introduction of additional phenomenological parameters (correlation length $\xi$,
and $\Delta$) through the definition of $\Sigma({\bf k},\omega)$ seems
to be a step back with respect the methods outlined above, it actually opens the
possibility to systematically distinguish between different types of
nonlocal fluctuations and their effects and helps 
to analyze experimental or theoretical data obtained
within more advanced schemes in terms of intuitive physical pictures. Note, however, that in principle 
even the paramters $\xi$ and $\Delta$ can be
calculated from the original model\cite{VT}, too.

An essential advantage of the proposed 
combination of two non-perturbative
methods (DMFT and $\Sigma({\bf k},\omega)$ from Refs.\cite{Sch,KS})
removes the restrictions on model parameters
in e.g.\ cluster mean-field theories.
Our scheme works for any Coulomb interaction strength $U$,
pseudogap strength $\Delta$, correlation length $\xi$, filling $n$
and bare electron dispersion $\varepsilon({\bf k})$ on a 2d square lattice
for any set of ${\bf k}$-points. Although we presented only
high-temperature
data in this paper, the possibility to use Wilson's NRG to solve the
effective impurity model also opens the possibiltiy to study
properties at $T=0$, which is currently impossible within the DCA or
CDMFT for larger clusters.
Moreover, the DMFT+$\Sigma_{\bf k}$ approach
can be easlily generalized to orbital degrees of freedom,
phonons, impurities, etc.

As a further application of our generalized DMFT+$\Sigma_{\bf k}$
we would like to bring readers attention to  Ref\cite{Kuchinskii05}, dealing
with the problem of Fermi surface destruction in High-T$_c$ compounds
because of pseudogap fluctuations.

\section{Acknowledgements}

We are grateful to A. Kampf for useful discussions.
This work was supported in part by RFBR grants 05-02-16301 (MS,EK,IN)
03-02-39024\_a (VA,IN), 04-02-16096 (VA,IN), 05-02-17244 (IN), by the joint 
UrO-SO project $No.$ 22 (VA,IN), and programs of the
Presidium of the Russian Academy of Sciences (RAS) ``Quantum macrophysics''
and of the Division of Physical Sciences of the RAS ``Strongly correlated
electrons in semiconductors, metals, superconductors and magnetic
materials''. I.N. acknowledges support
from the Dynasty Foundation and International
Centre for Fundamental Physics in Moscow program for young
scientists 2005), Russian Science Support Foundation program for
young PhD of Russian Academy of Science 2005. One of us (TP) further acknowledges
supercomputer support from the Norddeutsche Verbund f\"ur Hoch- und
H\"ochstleistungsrechnen.

\newpage


\appendix

\section{Derivation of generalized DMFT+$\Sigma_{\bf k}$ approach}
\label{A}

In this appendix we present a derivation of the generalized DMFT+$\Sigma_{\bf k}$ scheme
for the Hubbard model
\begin{equation}
H=-\sum_{ij,\sigma}{t_{ij}c_{i\sigma}^\dagger c_{j\sigma}}+
U\sum_i{n_{i\uparrow}n_{i\downarrow}}
\label{extended_Hubbard_model},
\end{equation}
using a diagrammatic approach.
The single--particle Green function in Matsubara representation is as usual given by
\begin{equation}
G_{\bf k}(i\omega)=\frac{1}{i\omega+\mu-\varepsilon({\bf k})
-\Sigma(i\omega,{\bf k})}
\label{GkH}
\end{equation}
To establish the standard DMFT one invokes the limit of infinite dimensions $d \to \infty$.
In this limit only local contributions to electron self--energy survive \cite{vollha93,georges96}, i.e.\ 
$\Sigma_{ij} \to \delta_{ij} \Sigma_{ii}$ or, in reciprocal space,
$\Sigma(i\omega,{\bf k}) \to \Sigma(i\omega)$.

\begin{figure}
\includegraphics[clip=true,width=0.8\textwidth]{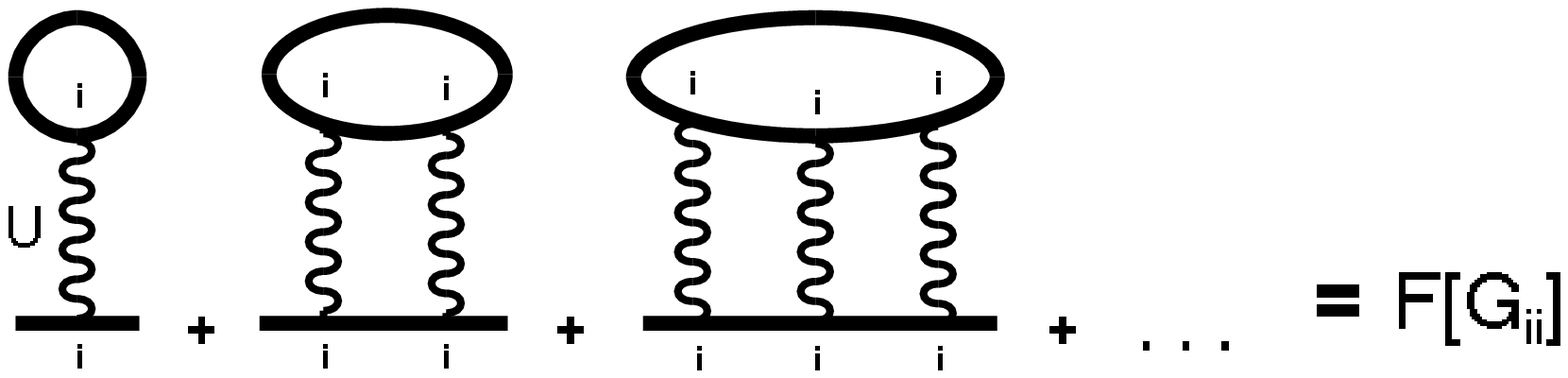}
\caption{Local ``skeleton'' diagrams for the DMFT self--energy $\Sigma$.
Wavy lines represent the local (Hubbard) Coulomb interaction $U$, full lines 
denote the local Green function $G_{ii}$.}
\label{sigmDMFT}
\end{figure}

In Fig. \ref{sigmDMFT} we show examples of ``skeleton'' diagrams for the local
self -- energy, contributing in the limit of $d \to \infty$. The complete series 
of these and similar diagrams defines the local self -- energy as a functional
of the local Green function
\begin{equation}
\Sigma = F[G_{ii}]\;,
\label{FGii}
\end{equation}
where
\begin{equation} 
G_{ii}(i\omega)=\frac{1}{N}\sum_{\bf k}\frac{1}{i\omega+\mu
-\varepsilon({\bf k})-\Sigma(i\omega)}.
\label{Gii}
\end{equation}
One then defines the ``Weiss field''
\begin{equation}
{\cal G}^{-1}_{0}(i\omega)=\Sigma(i\omega)+G^{-1}_{ii}(i\omega)
\label{Wssn}
\end{equation}
which is used to set up the effective single impurity problem with an effective action
given by (\ref{Seff}). Via Dyson's equation the Green function (\ref{AndImp})
for this effective single impurity problem can be written as
\begin{equation}
G_d(i\omega)=\frac{1}{{\cal G}^{-1}_{0}(i\omega)-\Sigma_d(i\omega)}
\label{Gd}
\end{equation}
and the ``skeleton'' diagrams for self--energy $\Sigma_d$ are just the same as
shown in Fig. \ref{sigmDMFT}, with the replacement $G_{ii} \to G_d$.
Thus we get
\begin{equation}
\Sigma_d = F[G_d],
\label{FGd}
\end{equation}
where $F$ is the same functional as in (\ref{FGii}). The two equations
(\ref{Gd}) and (\ref{FGd}) define both $G_d$ and $\Sigma_d$ for a given
``Weiss field'' ${\cal G}_0$. On the other hand, for the local
$\Sigma$ and $G_{ii}$ of the initial (Hubbard) problem we have precisely the
same  pair of equations, viz (\ref{FGii}) and (\ref{Wssn}), and ${\cal G}_0$ in both
problems is just the same, so that
\begin{equation}
\Sigma=\Sigma_d; \qquad G_{ii}=G_d.
\label{eqv}
\end{equation}
Thus, the task of finding the local self--energy of the 
$(d\to\infty)$ Hubbard model is eventually reduced to the calculation of the self--energy
of an effective quantum single impurity problem defined by effective action of Eq.
(\ref{Seff}).

\begin{figure}
\includegraphics[clip=true,width=0.8\textwidth]{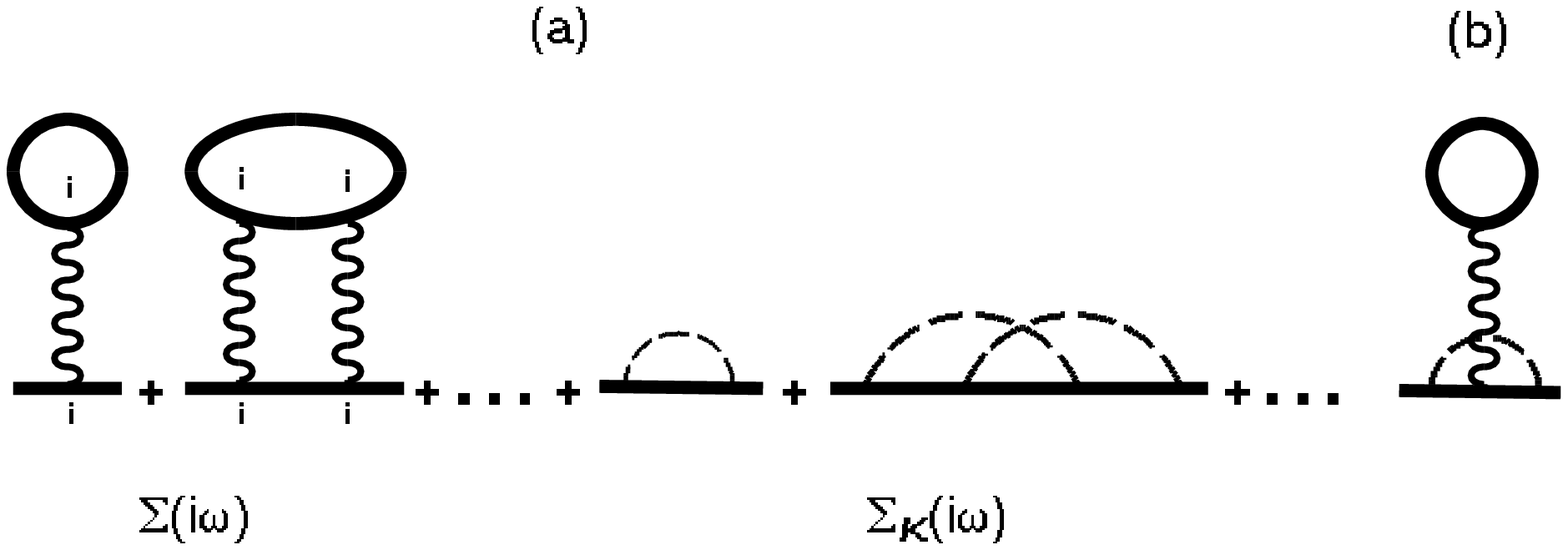}
\caption{Typical ``skeleton'' diagrams for the self--energy in the
DMFT+$\Sigma_{\bf k}$ approach.
The first two terms are DMFT self--energy diagrams; the
middle two diagrams show contributions to the
non-local part of the self--energy from
spin fluctuations (see section\protect\ref{kself}) represented
as dashed lines;
the last diagram (b) is an example of neglected diagram's leading to
interference
between the local and non-local parts. }
\label{dDMFT_PG}
\end{figure}

Consider now non -- local contribution to the self -- energy. If we neglect
interference between local and non--local contributions (as given e.g.
by the diagram shown in Fig.\ref{dDMFT_PG}(b)), the full self--energy is approximately
determined by the sum of these two contributions. ``Skeleton'' diagrams
for the non-local part of the self--energy, $\Sigma_{\bf k}(i\omega )$, 
are then those shown in Fig. \ref{dDMFT_PG}(a), where
the full line denotes the Green function $G_{\bf k}$ of Eq. (\ref{Gk}), 
while broken lines denote the interaction with static Gaussian spin (charge)
fluctuations.  These diagrams are just absent within the standard DMFT
(as any contribution from Ornstein -- Zernike type fluctuations vanish for
$d\to\infty$), and no double counting problems arise at all. 

The local contribution to the self--energy is again defined by the functional
(\ref{FGii}) via the local Green function $G_{ii}$, which is now given by
(\ref{Gloc}). Introducing again a ``Weiss field'' via (\ref{Wssn}) and repeating all
previous arguments, 
we again reduce the task of finding the local  part of the self--energy to
the solution of an ``single impurity'' problem with an effective action (\ref{Seff}).

To determine the non--local contribution $\Sigma_{\bf k}(i\omega )$ we first
introduce
\begin{equation} 
{\cal G}_{0\bf k}(i\omega)=\frac{1}{G_{\bf k}^{-1}(i\omega)+
\Sigma_{\bf k}(i\omega)}
=\frac{1}{i\omega+\mu-\varepsilon({\bf k})-\Sigma(i\omega)}
\label{G0pg}
\end{equation}
as the ``bare'' Green function for electron scattering by static Gaussian
spin (charge) fluctuations. The assumed static nature of these fluctuations
allows to use the method of Refs.\cite{Sch,KS,MS79} and the calculation of the 
non--local part of the self--energy $\Sigma_{\bf k}(i\omega )$ reduces to the recursion
procedure defined by Eqs.~(\ref{Sk}) and (\ref{rec}). The choice of the ``bare''
Green function Eq.~(\ref{G0pg}) guarantees that the Green function ``dressed'' 
by fluctuations
$G_{\bf k}^{-1}(i\omega)={\cal G}_{0\bf k}^{-1}(i\omega)-\Sigma_{\bf k}(i\omega)$, 
which enters into the ``skeleton'' diagrams for $\Sigma_{\bf k}(i\omega)$, just
coincides with the full Green functions $G_{\bf k}(i\omega)$.

Thus we obtain a fully self--consistent scheme to calculate both local
(due to strong single--site correlations) and non--local (due to short--range
fluctuations) contributions to electron self--energy.

\section{$\Delta$ in the Hubbard model.}
\label{B}

In this Appendix we derive the explicit microscopic expression for
pseudogap amplitude $\Delta$ given in (\ref{DeltHubb}). Within the two--particle
self--consistent approach of Ref.~\cite{VT}, valid for medium values of
$U$,  and neglecting charge fluctuations, we can write down an expression for the
electron self--energy of the form used in (\ref{Gk}), with
\begin{equation}
\Sigma_{\sigma}(i\omega)=Un_{-\sigma}
\label{Sigloc}
\end{equation}
as the lowest order local contribution due to the on--site Hubbard interaction,
surviving in the limit of $d\to\infty$, and exactly accounted for in DMFT
(with all higher--order contributions).
Non--local contribution to the self--energy (vanishing for $d\to\infty$ and
not accounted within DMFT) due to interaction with spin--fluctuations then 
leads to the expression
\begin{equation}
\Sigma_{\vec k}(i\omega)=\frac{U}{4}\frac{T}{N}\sum_m\sum_{\bf q}
U_{sp}\chi_{sp}({\bf q},\nu_m)G_0({{\bf{k+q}},i\omega+i\nu_m})\;,
\label{Signloc}
\end{equation}
where
\begin{equation}
U_{sp}=g_{\uparrow\downarrow}(0)U,\qquad
g_{\uparrow\downarrow}(0)=\frac{<n_{i\uparrow}n_{i\downarrow}>}
{<n_{i\uparrow}><n_{i\downarrow}>}
\label{Usp}
\end{equation}
with $<n^2_{\sigma}>=<n_{\sigma}>$ and
$<n_{i\uparrow}>=<n_{i\downarrow}>=\frac{1}{2}n$ in the paramagnetic phase.
For the dynamic spin susceptibility $\chi_{sp}({\bf q},\nu_m)$ we use
the standard Ornstein--Zernike form \cite{VT}, similar to that
used in spin--fermion model \cite{Sch}, which describes enhanced
scattering with momenta transfer close to antiferromagnetic vector
${\bf Q}=(\pi/a,\pi/a)$. With these approximations, we can write down the following
expression for the non--local contribution to the self--energy \cite{Sch,KS}:
\begin{eqnarray}
\Sigma_{\vec k}(i\omega)=\frac{1}{4}UU_{sp}\frac{T}{N}\sum_m\sum_{\vec q}
\chi_{sp}({\bf q},\nu_m)\frac{1}{i\omega+i\nu_m+\mu-\varepsilon({\bf k+q})}
\approx\nonumber\\
\approx\frac{1}{4}UU_{sp}\frac{T}{N}\sum_m\sum_{\vec q}\chi_{sp}({\bf q},\nu_m)
\sum_{\vec q}S({\vec q})\frac{1}{i\omega+\mu-\varepsilon({\bf k+q})}
\equiv\nonumber\\
\equiv\Delta^2\sum_{\vec q}S({\bf q})\frac{1}{
i\omega+\mu-\varepsilon({\bf k+q})}=\nonumber\\
=\frac{\Delta^2}{i\omega+\mu-\varepsilon({\bf p+Q})
+i(|v^x_{\bf p+Q}|+|v^y_{\bf p+Q}|)\kappa {\rm sign}\omega}.
\label{SigKS}
\end{eqnarray}
Here we have introduced the static form factor \cite{KS}
\begin{equation}
S({\vec q})=\frac{2\xi^{-1}}{(q_x-Q_x)^2+\xi^{-2}}
\frac{2\xi^{-1}}{(q_y-Q_y)^2+\xi^{-2}}
\label{Sq}
\end{equation}
and the squared pseudogap amplitude
\begin{eqnarray}
\Delta^2=\frac{1}{4}UU_{sp}\frac{T}{N}
\sum_m\sum_{\vec q}\chi_{sp}({\bf q},\nu_m)=\nonumber\\
=\frac{1}{4}UU_{sp}[<n_{i\uparrow}>+<n_{i\downarrow}>
-2<n_{i\uparrow}n_{i\downarrow}>]=\nonumber\\
=\frac{1}{4}UU_{sp}\frac{1}{3}<{\vec S}_i^2>,
\label{Delta2}
\end{eqnarray}
where we have used the exact sum--rule for the susceptibility \cite{Sch,VT}.
Taking into account (\ref{Usp}) we immediately obtain (\ref{DeltHubb}).

Actually, the approximations made in (\ref{SigKS}) and (\ref{Sq}) allow for an exact
summation of the whole Feynman series for electron interaction with spin--fluctuations,
replaced by the static Gaussian random field.
Thus generalizing the one--loop approximation (\ref{SigKS}) eventually leads to the
basic recursion procedure given in (\ref{Sk}), (\ref{rec}) Refs.~\cite{Sch,KS}.

\begin{figure}
\includegraphics[clip=true,width=0.8\textwidth]{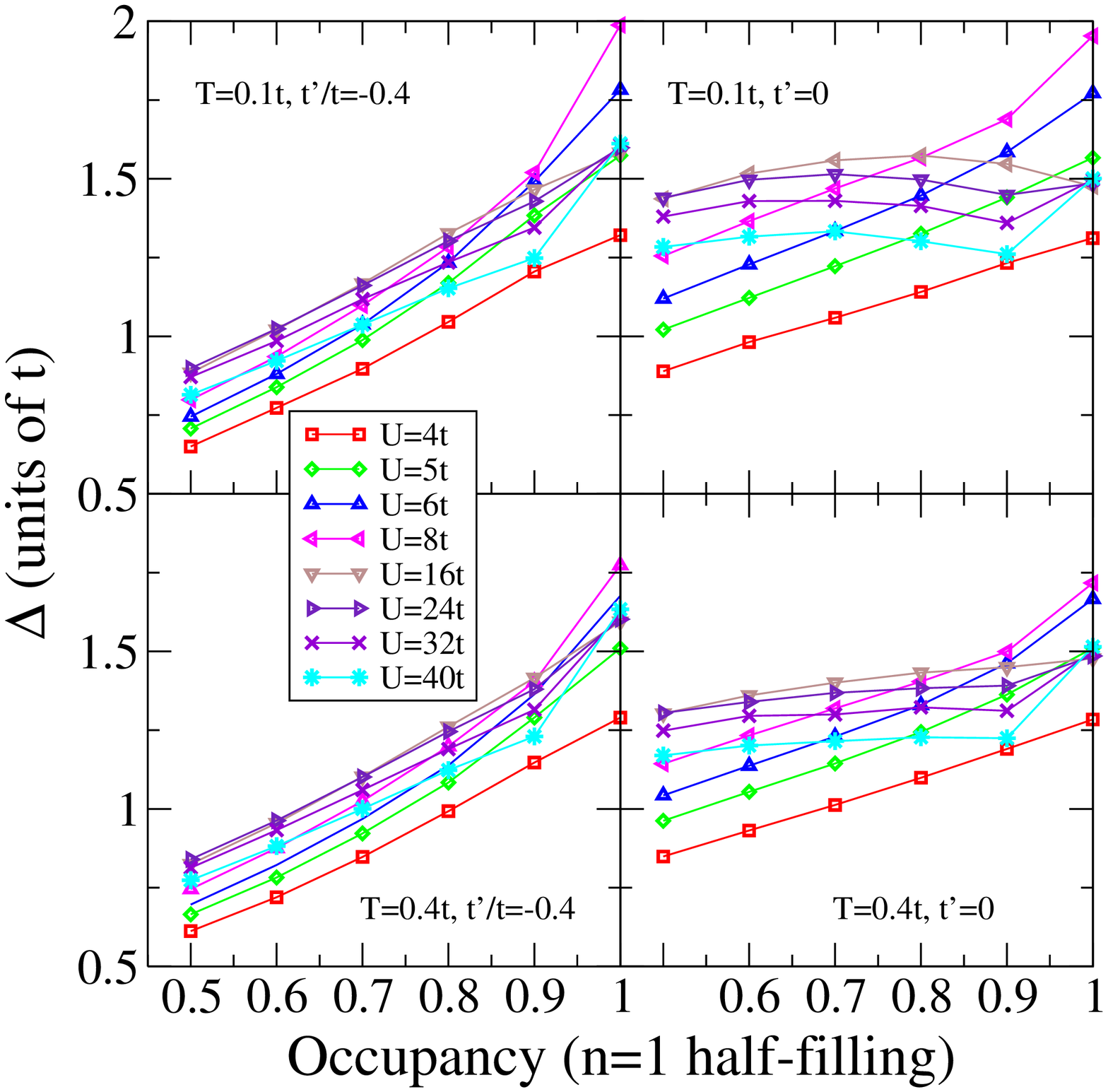}
\caption{(Color online) Filling dependence of the pseudo gap potential $\Delta$
calculated with DMFT(QMC)  
for varying Coulomb interaction ($U$) and temperature ($T$)
on a two--dimensional square lattice with two sets of ($t,t'$).}
\label{delta}
\end{figure}

Using the DMFT(QMC) approach we computed occupancies $<n_{i\uparrow}>$, $<n_{i\downarrow}>$
and double occupancies $<n_{i\uparrow}n_{i\downarrow}>$ required to
calculate the pseudogap amplitude $\Delta$ of Eq. (\ref{Delta2})
In Fig. \ref{delta} the corresponding values of $\Delta$ are presented.
One can see that $\Delta$ grows when the filling goes to $n=1$.
While $U$ approaches $8t$ (the value of the bandwidth for a square lattice)
$\Delta$ as a function of $n$ grows monotonically. When $U$ becomes larger
than $W=8t$ (when a metal--insulator transition occurs) 
one can see a local minimum for 
$n=0.9$, which becomes more pronounced with further increase of $U$.
For $t'/t=-0.4$ and both temperatures the scatter of $\Delta$ values is 
smaller than for the case of $t'=0$. Also $\Delta$ has a rather weak 
temperature dependence.  
All values of $\Delta$ lie in the interval $\sim 
0.75t \div 2t$.  Therefore, for our computations we took only two  
characteristic values of $\Delta=t$ and $\Delta=2t$.

\pagestyle{empty}

\newpage

\newpage

\newpage

\newpage

\newpage

\newpage

\newpage

\newpage

\newpage

\newpage

\newpage

\newpage

\begin{thebibliography}{99}
\vfill

\bibitem{Tim}T. Timusk, B. Statt, Rep. Progr. Phys, {\bf 62}, 61 (1999).

\bibitem{MS}M. V. Sadovskii, Usp. Fiz. Nauk {\bf 171}, 539 (2001)
[Physics -- Uspekhi {\bf 44}, 515 (2001)].

\bibitem{Pines}D. Pines,\ ArXiv:\ cond-mat/0404151.

\bibitem{Sch}J. Schmalian, D. Pines, B.Stojkovic, Phys. Rev. Lett. {\bf 80}, 3839 (1998);
Phys. Rev. B {\bf 60}, 667 (1999).

\bibitem{KS}E. Z. Kuchinskii, M. V. Sadovskii, Zh. Eksp. Teor. Fiz. {\bf 115},
1765 (1999) [(JETP {\bf 88}, 347 (1999)].
(available as ArXiv: cond-mat/9808321)

\bibitem{MetzVoll89}  W. Metzner and D. Vollhardt, Phys. Rev. Lett.
{\bf 62}, 324 (1989).

\bibitem{vollha93}  D.~Vollhardt, in {\em Correlated Electron Systems},
edited by V.~J. Emery, World Scientific, Singapore, 1993, p.~57.

\bibitem{pruschke}  Th. Pruschke, M. Jarrell, and J. K. Freericks, Adv.
in Phys. {\bf 44}, 187 (1995).

\bibitem{georges96}  A. Georges, G. Kotliar, W. Krauth, and M. J. Rozenberg,
Rev. Mod. Phys. {\bf 68}, 13 (1996).

\bibitem{PT} G.\ Kotliar and D.\ Vollhardt, Physics Today \textbf{57},
No.\ 3 (March), 53 (2004).

\bibitem{Si96}Q. Si and J.L. Smith, Phys. Rev. Lett. {\bf 77}, 3391 (1996).

\bibitem{TMrmp} Th.\ Maier, M.\ Jarrell, Th.\ Pruschke and M.\ Hettler,
Rev.\ Mod.\ Phys.\ (in print, ArXiv: cond-mat/0404055).

\bibitem{KSPB}G. Kotliar, S.Y. Savrasov, G. Palsson, G. Biroli,
Phys. Rev. Lett. {\bf 87}, 186401 (2001); For periodized version (PCDMFT)
see M. Capone, M. Civelli, S.S. Kancharla, C. Castellani, and G. Kotliar,
Phys.\ Rev.\ B {\bf 69}, 195105 (2004).

\bibitem{Gross94} C.~Gros, and R.~Valenti, Annalen der Phys. {\bf 3},
  460 (1994).

\bibitem{Senechal00}D. Senechal, D. Perez, and M. Pioro-Ladrière,
Phys. Rev. Lett. {\bf 84}, 522 (2000);
D. Senechal, D. Perez, and D. Plouffe,
Phys.\ Rev.\ B {\bf 66}, 075129 (2002).

\bibitem{Kyung05}B. Kyung, S.S. Kancharla, D. Senechal, A.-M.S.
Tremblay, M. Civelli, G. Kotliar, ArXiv: cond-mat/0502565.

\bibitem{CivKot}M. Civelli, M. Capone, S.S. Kancharla, O. Parcollet,
G. Kotliar, ArXiv:\ cond-mat/0411696.

\bibitem{QMC}  J. E. Hirsch and R. M. Fye, Phys. Rev.
Lett. {\bf 56}, 2521
(1986); M. Jarrell, Phys. Rev. Lett. {\bf 69}, 168 (1992); M. Rozenberg,
X. Y. Zhang, and G. Kotliar, Phys. Rev. Lett. {\bf 69}, 1236 (1992);
A. Georges and W. Krauth, Phys. Rev. Lett. {\bf 69}, 1240 (1992);
M. Jarrell in {\it{Numerical Methods for lattice Quantum Many-Body Problems}},
edited by D. Scalapino, Addison Wesley, 1997. For review of QMC for DMFT
see~Ref.\protect\cite{Held01}.

\bibitem{Held01} K. Held, I.A. Nekrasov, N. Bl\"umer, V.I. Anisimov, and D. Vollhardt,
Int. J. Mod. Phys. B {\bf 15}, 2611 (2001);  K. Held, I.A. Nekrasov, G. Keller,
V. Eyert, N. Bl\"umer, A.K. McMahan, R.T. Scalettar, T. Pruschke,
V.I. Anisimov, and D. Vollhardt, ArXiv: cond-mat/0112079
(Published in {\it Quantum Simulations of
Complex Many-Body Systems: From Theory to Algorithms},
eds. J.~Grotendorst, D.~Marks, and A.~Muramatsu, NIC Series Volume 10
(NIC Directors, Forschunszentrum J\"ulich, 2002) p. 175-209.

\bibitem{NRG} K.G.~Wilson,  Rev.\ Mod.\ Phys. {\bf 47}, 773 (1975);
  H.R.~Krishna-murthy, J.W.~Wilkins, and K.G.~Wilson,
  Phys.\ Rev.\ B {\bf 21}, 1003 (1980); {\it ibid.} {\bf 21}, 1044 (1980);
for a comprehensive introduction to teh NRG see e.g.\  A.C.\ Hewson,
  {\em The Kondo Problem to Heavy Fermions} (Cambridge University Press,
  1993).
\bibitem{BPH}R.~Bulla,  A.C.~Hewson and Th.~Pruschke,
  J.\ Phys.\ -- Condens.\ Matter {\bf 10}, 8365(1998); 
R.~Bulla, Phys.\ Rev.\ Lett. {\bf 83}, 136 (1999).

\bibitem{MS79}M. V. Sadovskii, Zh. Eksp. Teor. Fiz. {\bf 77}, 2070(1979)
[Sov.Phys.--JETP {\bf 50}, 989 (1979)].

\bibitem{VT}Y. M. Vilk, A.-M. S. Tremblay, J. Phys. I France {\bf 7}, 1309 (1997).

\bibitem{Gunnarsson89}O. Gunnarsson, O. K. Andersen, O. Jepsen, and J. Zaanen,
Phys. Rev. B \textbf{39}, 1708 (1989).

\bibitem{LSCO_U_value}M. T. Czyzyk and G. A. Sawatzky, Phys. Rev. B{\bf 49}, 14211 (1994).

\bibitem{cm05}M.V. Sadovskii, I.A. Nekrasov, E.Z. Kuchinskii, Th. Prushke,
V.I. Anisimov. ArXiv: cond-mat/0502612.

\bibitem{mpj03} Th.A.~Maier, Th.~Pruschke, and M.~Jarrell, Phys.\
  Rev.\ B {\bf 66}, 075102 (2002).

\bibitem{Stanescu03} T.D. Stanescu and P. Phillips,
 Phys. Rev. Lett. {\bf 91}, 017002 (2003). 

\bibitem{Katanin04}A.A. Katanin and A.P. Kampf, Phys. Rev. Lett. {\bf 93}, 106406 (2004) 

\bibitem{RM}D. Rohe, W. Metzner, Phys. Rev. B{\bf 71}, 115116 (2005).

\bibitem{Barisic05}D.K. Sunko, S. Barisic, Eur. Phys. J. B {\bf 46}, 269 (2005).

\bibitem{Kam}A. Kaminski, H. M. Fretwell, M. R. Norman, M. Randeria, S. Rosenkranz,
U. Chatterjee, J. C. Campuzano, J. Mesot, T. Sato, T. Takahashi, T. Terashima,
M. Takano, K. Kadowaki, Z. Z. Li, H. Raffy, Phys.\ Rev.\ B {\bf 71}, 014517 (2005).

\bibitem{Senechal05}D. Senechal and A.-M.S. Tremblay,
Phys. Rev. Lett. {\bf 92}, 126401 (2004).

\bibitem{HW}K. Haule, A. Rosch, J. Kroha, P. W\"olfle, Phys. Rev. Lett.
{\bf 89}, 236402 (2002);\ Phys. Rev. B {\bf 68},
155119 (2003).

\bibitem{KyTr}B. Kyung, V. Hankevich, A.-M. Dare, A.-M.S. Tremblay.
Phys. Rev. Lett. {\bf 93}, 147004 (2004). 

\bibitem{Prelovsek}P. Prelovsek and A. Ramsak, Phys. Rev. B {\bf 63}, 180506 (2001);
P. Prelovsek, A. Ramsak, ArXiv: cond-mat/0502044.

\bibitem{BAG}S. Biermann, F. Aryasetiawan, A. Georges, Phys. Rev. Lett.
{\bf 90}, 086402 (2003).

\bibitem{SKot}P. Sun, G. Kotliar, Phys. Rev. Lett. {\bf 92}, 196402 (2004).

\bibitem{Kuchinskii05}E.Z. Kuchinskii, I.A. Nekrasov, M.V. Sadovskii,
JETP Letters {\bf 82}(4), 217 (2005).

\end{thebibliography}
\end{document}